# Moiré-Driven Interfacial Thermal Transport in Twisted Transition Metal Dichalcogenides

Wenwu Jiang[1], Ting Liang[2], Hekai Bu[1], Jianbin Xu[2], Wengen Ouyang[1,3*]

[1] Department of Engineering Mechanics, School of Civil Engineering, Wuhan University, Wuhan, Hubei 430072, China

[2] Department of Electronic Engineering and Materials Science and Technology Research Center, The Chinese University of Hong Kong, Shatin, N.T., Hong Kong SAR, 999077, P. R. China

[3] State Key Laboratory of Water Resources Engineering and Management, Wuhan University, Wuhan, Hubei 430072, China

[*]E-mail: w.g.ouyang@whu.edu.cn (W. Ouyang)

**Keywords:** moiré superlattice; twist engineering; atomic reconstruction; transition metal dichalcogenides; interfacial thermal transport; registry-dependent interlayer potential

**Abstract**

Cross-plane thermal conductivity in homogeneous transition metal dichalcogenides (TMDs) exhibits a pronounced dependence on interfacial twist angle, originating from atomic reconstruction within moiré superlattices. This reconstruction redistributes interlayer stacking modes, reducing high-efficiency thermal transport regions and softening the transverse acoustic phonon modes as the twist angle increases. We propose a general theoretical framework to capture this behavior, validated against non-equilibrium molecular dynamics simulations across both homo- and heterogeneous twisted TMDs structures, as well as homogeneous twisted graphene and hexagonal boron nitride stacks. Our model reveals that the interfacial thermal conductance (ITC) scales with the twist angle ($\theta$) as $\ln(\mathrm{ITC}) \propto e^{-\sqrt{\theta}}$. These findings advance the understanding of twist-engineered interfacial thermal transport, offering design principles for optimizing thermal management in devices based on van der Waals layered materials.

Moiré superlattices formed from interlayer twisting or lattice mismatch of van der Waals (vdW) layered materials, such as graphene,[1-3] hexagonal boron nitride ($h$-BN),[4] and transition metal dichalcogenides (TMDs),[5-7] have attracted significant scientific and technological attention due to their unique electronic, optical and tribological characteristics, including exceptional moiré excitons,[8-11] unconventional superconductivity,[12] ferromagnetism and ferroelectricity,[13-16] and superlubricity.[17-19] Recently, "twisted thermotics"[20] has emerged as a compelling field, focusing on the twist angle dependence of in-plane and cross-plane thermal transport in twisted vdW layered materials,[21-24] including twisted $MoS_2$ and $WS_2$,[25, 26] twisted multilayer graphene and $h$-BN[27, 28] and other twisted vdW interfaces.[29-31] Especially, Cheng *et al.*[24] found there exists a "magic angle" in the in-plane thermal conductivity of twisted bilayer graphene. While substantial progress has been made in understanding and manipulating thermal conductivities in these structures, the explanations have primarily relied on complex phonon transport concepts. A direct, analytical relationship between cross-plane thermal conductivity and moiré superlattices remains elusive, often necessitating costly experiments or computationally intensive simulations.

To address this issue, we systematically investigated the interfacial thermal transport properties of twisted 2H-TMD stacks. Our findings reveal that the twist-angle dependence of cross-plane thermal conductivity originates from atomic reconstruction within moiré superlattices. This reconstruction redistributes interlayer stacking modes, reducing high-efficiency thermal transport regions and softening the transverse acoustic phonon modes as the twist angle increases. Based on these insights, we establish a direct analytical relation between the interfacial thermal conductance (ITC) and the twist angle ($\theta$), which scales as $\ln(\mathrm{ITC}) \propto e^{-\sqrt{\theta}}$. This concise expression provides a framework for elucidating and modulating the thermal transport properties of twisted vdW interfaces. Remarkably, the ITC predicted by this analytical equation shows excellent agreement with our non-equilibrium molecular dynamics (NEMD) simulation results for various twisted vdW interfaces. Our study thus provides a powerful predictive tool to accelerate progress in thermal management of van der Waals layered materials and offers a promising avenue for exploring fundamental physical phenomena in these systems.

## RESULTS AND DISCUSSION

We employ NEMD simulations with the state-of-the-art force fields to explore interfacial thermal transport in homogeneous 2H-TMD stacks ($MoS_2$, $MoSe_2$, $WS_2$, and $WSe_2$) across a wide range of twist angles, considering both periodically parallel (P) and antiparallel (AP) stacking configurations. The details of building periodic commensurate structures are presented in the Methods section. Figure 1a depicts the NEMD simulation setup for calculating the cross-plane thermal conductivity of the twisted TMDs. Previous studies[28, 29, 32, 33] have demonstrated that this NEMD simulation setup can successfully establish a temperature gradient and lead to reasonable cross-plane thermal conductivities. The Stillinger-Weber (SW)[34] potential and the anisotropic interlayer potential (ILP)[35, 36] were used to describe the intralayer interactions within each layer and the interlayer interaction between TMDs layers, respectively. The applicability of ILP to vdW layered materials, along with its accuracy for describing their mechanical, thermal transport and tribological properties has recently been demonstrated.[19, 28, 29, 37-46] These features, however, cannot be captured with the commonly adopted pairwise Lennard-Jones (LJ)[47] potential.[28, 36, 40, 41, 46, 48-52] Our benchmark test [see Section 1.1-1.2 of the Supporting Information (SI)] shows that the isotropic LJ potential fails to reproduce the atomic reconstruction of moiré superlattices and phonon spectra of TMD systems, whereas the ILP achieves excellent agreement with experimental observations. Moreover, the cross-plane thermal conductivity of bulk AA′-stacked $MoS_2$ calculated using the SW+ILP potential within the homogeneous nonequilibrium molecular dynamics (HNEMD)[53] method, as implemented in GPUMD package,[54] are in good agreement with experimental and DFT results,[25, 30, 55-61] confirming the accuracy of this approach (SW+ILP) for thermal transport calculations (see Section 1.3 of SI). These benchmark tests thus further strongly affirm the validity of ILP in accurately describing the interlayer interaction in twisted TMDs interfaces. The details of NEMD simulations and convergence tests are presented in the Methods section and Sections 2 and 3 of the SI, respectively.

Figure 1b shows the cross-plane thermal conductivity ($\kappa_{\mathrm{CP}}$) (see computational details of $\kappa_{\mathrm{CP}}$ in Methods section) of the entire stacks as a function of the twist angle for both parallel and antiparallel configurations ($MoSe_2$ and $WS_2$), each consisting of 16 layers of TMDs (the results of $MoS_2$ and $WSe_2$ are shown in Section 4 of SI). A significant dependence of $\kappa_{\mathrm{CP}}$ on twist angle is observed in all four TMD systems, and the $\kappa_{\mathrm{CP}}$ decreases quickly in the range of 0° ~ 5° with overall 70~80% reductions. Similar results have been observed for multilayer twisted graphene, *h*-BN, and $MoS_2$ systems.[26, 28, 29] Notably, the experimental value of $\kappa_{\mathrm{CP}}$ (0.041 ± 0.003 $\mathrm{Wm^{-1}K^{-1}}$)[25] for randomly twisted $WS_2$ falls within the range of our simulations, which indicates the accuracy of the anisotropic ILP for describing the vdW interaction across the twisted multilayer TMDs stacks. To further understand the impact of twist engineering on interfacial thermal transport, we investigated the ITR,[28, 29, 62] which quantifies the thermal transport resistance between two adjacent layers at the twisted interface (see computational details of ITR in Method section). Figure 1c shows a pronounced twist-angle dependence of ITR at the twisted interface, clearly demonstrating the feasibility of twist engineering in regulating the thermal transport properties of layered materials. Thermal conductivity at the nanoscale is known to be highly sensitive to sample dimensions. To systematically explore this size dependence, we performed a series of simulations by varying both the thickness and in-plane dimensions of the twisted structures. Our results reveal that $\kappa_{CP}$ increases with thickness, a behavior driven by the suppression of low-frequency phonons (with long mean free paths) due to the finite length along the heat flux direction.[57, 63, 64] Furthermore, we found that when the lateral cross-sectional size exceeds a critical threshold, its effect on interlayer thermal conductivity becomes negligible (seev Section 3.1 of the SI). While system thickness influences the absolute cross-plane thermal conductivity, the normalized data reveals a consistent twist-angle dependence regardless of thickness (see Section 3.2 of the SI). This consistency demonstrates that the fundamental mechanism underlying twist-angle modulation of interlayer thermal conductivity is not dominated by the sample size.

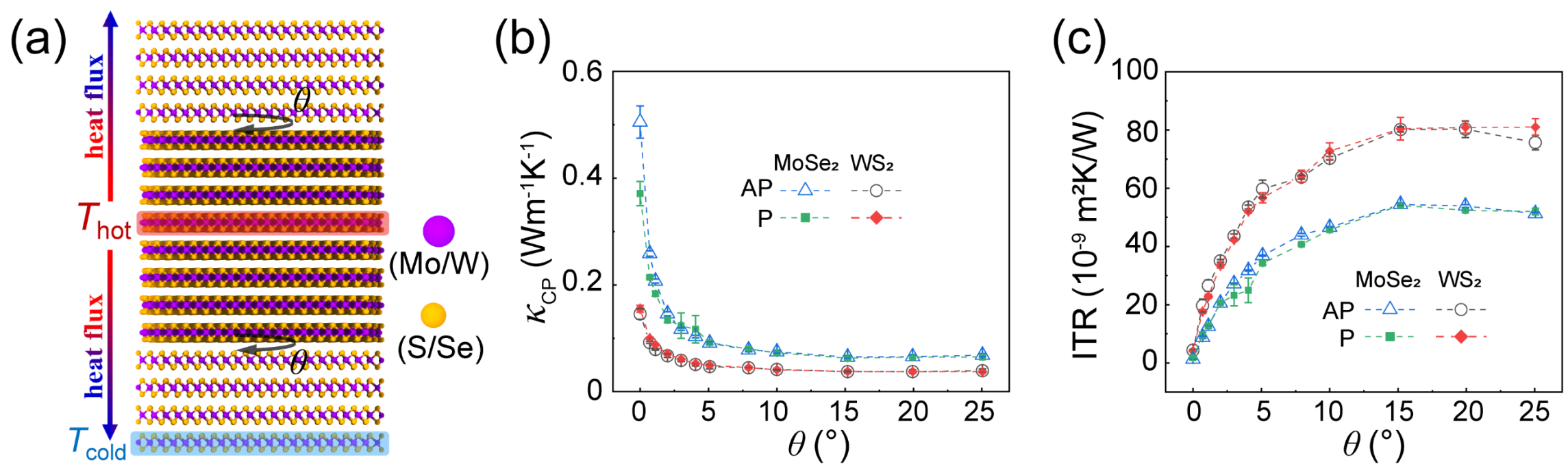

Figure 1. Twist engineering of interfacial thermal transport properties of 2H-TMD structures. (a) The schematic model of simulation setup. Two identical AB (AA′)-stacked TMD slabs for parallel (antiparallel) configurations are twisted with respect to each other to create a stacking fault of misfit angle $\theta$. Purple (yellow) spheres correspond to metal (chalcogen) atoms. A thermal bias is induced by applying Langevin thermostats to the two layers marked by red (heat source) and blue (heat sink) rectangles. The arrows indicate the direction of the vertical heat flux. Periodic boundary conditions are applied in the vertical direction, resulting in two twisted interfaces, and heat flows are in opposite directions across these interfaces. (b) Twist-angle dependence of cross-plane thermal conductivity ($\kappa_{CP}$) of the entire stacks. (c) Interfacial thermal resistance (ITR) of the twisted contact formed between the optimally stacked slabs of TMDs.

To elucidate the fundamental mechanism by which the twist angle modulates thermal transport between TMD layers, we conducted a comprehensive investigation into the intrinsic thermal properties of the moiré superlattices formed at the twisted interface. In typical twisted TMD stacks, different regions of the moiré superlattices exhibit varying stacking modes due to atomic reconstruction. As shown in Figures 2a,b, the twisted interface of the 16-layer $WS_2$ stacks with a twist angle of $\theta = 0.25°$ reveals four high-symmetry stacking configurations: AA, AB, BA, and domain wall (DW) for the parallel arrangement (see Figure 2c), and AA′, AB′, A′B, and DW for the antiparallel arrangement (see Figure 2d).

To investigate the effect of stacking modes within moiré superlattices on the interfacial thermal transport properties of the system, we first computed the interlayer binding energy using ILP to understand the energetics of domain formation (see Section 1 of the SI). For parallel configurations, we found that the most energetically favorable stackings are AB and BA, which are symmetric with respect to the basal plane reflection and have the same energy. In contrast, for antiparallel configurations, the stacking modes are energetically favored in the order of AA′ > AB′ > DW > A′B. This energetic preference is reflected in the disparity of domain sizes observed in Figure 2b. Subsequently, we calculated the ITR for various ideal high-symmetry stacks to investigate the relationship between ILP binding energy and interfacial thermal transport. As shown in Figure 2e, the normalized ITR of the homogeneous interface strongly depends on the stacking models with different ILP energies. Specifically, stacking with lower binding energies (absolute value) exhibits larger ITR. This observation is further corroborated by calculating the ILP binding energy as a function of twist angle $\theta$. As shown in Figure 2f, the absolute value of the ILP energy continuously increases and stabilizes as $\theta$ increases, which is consistent with the aforementioned dependence of ITR across twisted TMD interface on the twist angle (the ILP energy surfaces are presented in Section 5 of SI). Moreover, regulating the ILP energy while leaving the intralayer interaction unchanged is sufficient to achieve a dramatic shift in the cross-plane thermal conductivity. As is presented in Section 6 of SI, the $\kappa_{\mathrm{CP}}$ increases almost linearly with increasing the strength of ILP for both the parallel and antiparallel TMD structures. This relationship provides strong evidence that the enhancement of ILP strength leads to stronger phonon-phonon coupling at the interface, resulting in higher interfacial thermal conductivity.

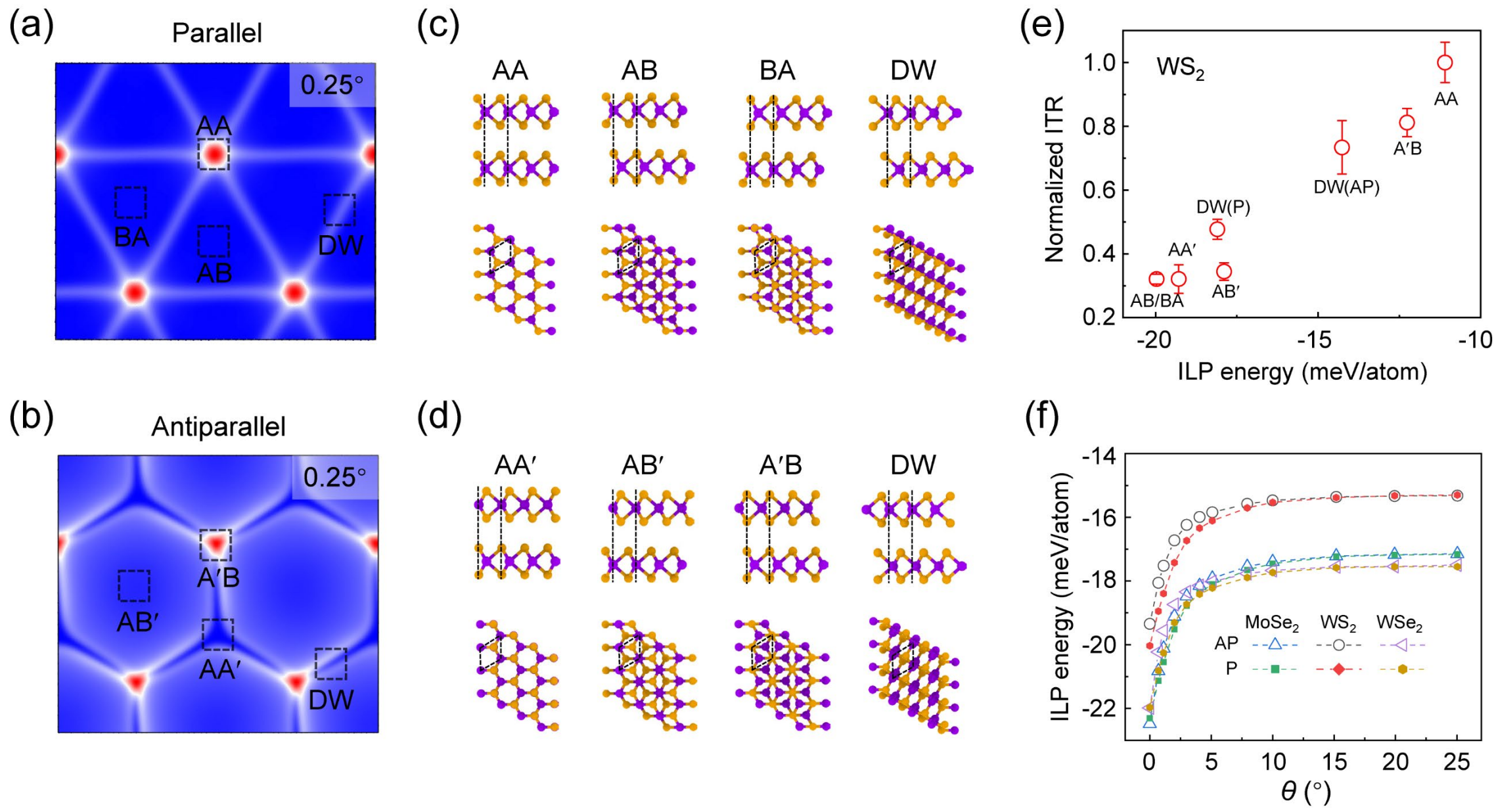

Figure 2. Effect of stacking modes and binding energy on thermal transport properties. Local registry index (LRI)[65, 66] corrugations of twisted $WS_2$ bilayers with $\theta = 0.25°$ in the parallel configuration (a) and the antiparallel configuration (b), relaxed using SW+ILP. Four high-symmetry stacking modes for parallel alignment (a, c: AA, AB, BA, and DW) and antiparallel alignment (b, d: AA′, AB′, A′B, and DW) of twisted bilayer $WS_2$ with a twist angle of $\theta = 0.25°$. Purple (yellow) spheres correspond to metal (chalcogen) atoms. (e) The ILP energy dependence of the ITR, which is normalized by the ITR of AA stacking for the ideal high-symmetry stacking modes. (f) Twist-angle dependence of ILP potential energy of the twisted contact formed between the optimally stacked slabs of TMDs.

To further qualitatively investigate the effect of stacking modes on the thermal transport properties, we conducted an in-depth analysis of different high-symmetry stackings in the twisted interface using spectral heat current (SHC)[29, 67, 68] decomposition methods (see Method section). Figure 3a presents the frequency-dependent spectral phonon thermal conductance $G(\omega)$ of various high-symmetry stacking regions in the twisted interface of parallel $WS_2$ system with $\theta = 1.12°$. The results demonstrate that interlayer thermal transport in regions with different stacking modes is dominated by phonons at distinct frequency ranges, and indicate distinct thermal transport capacities among the local stacking arrangements (the method for separating these stacking modes can be found in Section 7 of the SI), which can be ranked as follows: AB/BA > DW > AA. This observation is further supported by the calculation of the ITC, which represents the cumulative effect of all phonon eigenmodes across a broad spectrum of frequencies for four typical stackings in the twisted interface, as determined using the SHC method (see Figure 3b). For antiparallel $WS_2$ configuration, similar behavior has been observed. Based on calculated thermal properties (see Figures 3c,d), the thermal transport capacities for different stacking modes can be ranked as follows: AA′ > AB′ > DW > A′B . This hierarchy in thermal transport properties for both parallel and antiparallel configurations aligns with our findings for ideal high-symmetry stacks, as illustrated in Figure 2e. Furthermore, we extended our investigation to other 2H-TMD systems, namely $MoSe_2$ and $WSe_2$. The results, detailed in Section 7 of the SI, demonstrate similar trends in thermal transport properties across different stacking configurations. This consistency across various 2H-TMD materials suggests that our findings may be applicable to a broader range of layered materials, highlighting the potential for generalizing these insights in the field of 2D material thermal management.

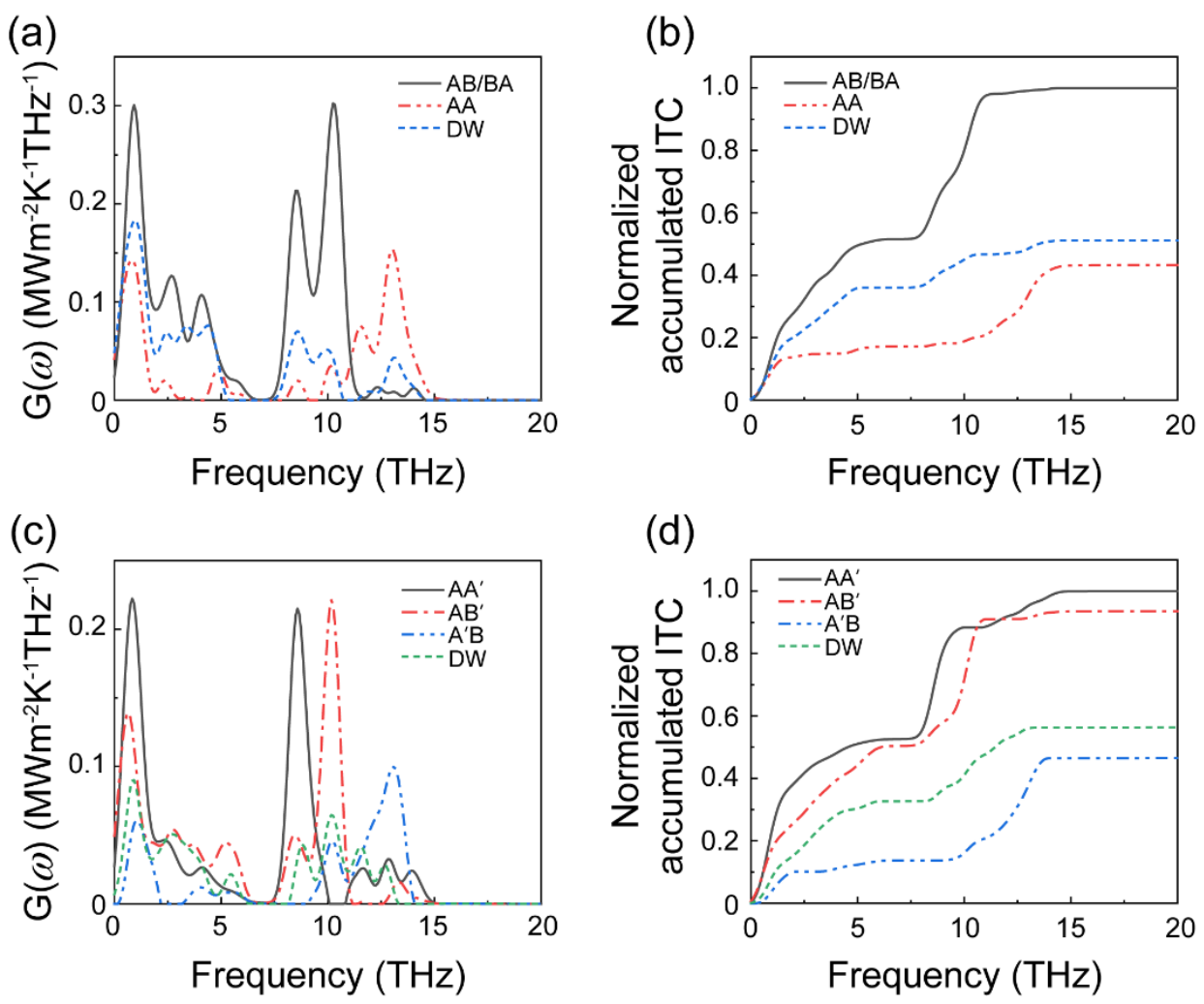

Figure 3. Thermal transport characteristics of different high-symmetry stacking modes in the twisted interface. Spectral phonon thermal conductance $G(\omega)$ across the twisted interface is calculated by the spectral heat current method for four high-symmetry stacking modes of parallel alignment (a) and antiparallel alignment (c). (b, d) Accumulated ITC normalized by the ITC of AB/BA stacking for parallel alignment (b) and AA' stacking for antiparallel alignment (d).

Based on the results presented above, we defined AB/BA (AA′/AB′) stackings of parallel (antiparallel) configurations as high-efficiency thermal transport regions, while categorizing AA and DW (A′B and DW) stackings as low-efficiency thermal transport regions. Then we employed the LRI[65, 66] method, which is suitable for characterizing surface reconstruction in twisted TMD interfaces to identify the twist-angle dependence of the stacking arrangement distribution. Figure 4a clearly shows that the distribution of high-efficiency thermal transport region (blue area) in both parallel and antiparallel configurations gradually decreases with increasing twist angle, mirroring the twist-angle dependence of $\kappa_{\mathrm{CP}}$. To quantify the size of these regions, we calculated their proportion in the twisted system using LRI (see Section 8 of the SI). Figures 4b,c reveal that the proportion of high-efficiency thermal transport region is significantly higher at small twist angles (0° ~ 5°) compared to that at larger twist angles (> 5°). When TMD layers are twisted at small angles, a large-scale moiré pattern emerges at the interface. Before reconstruction, the proportion of high-efficiency thermal transport regions (blue triangles) remains relatively constant at approximately 2% ~ 5% for all non-zero twist angles $\theta$. Upon relaxation, this pattern evolves into a periodic superstructure with clear domains. In this process, the upper and lower layers undergo slight in-plane deformation (atomic reconstruction) to achieve near commensuration, which involves bond stretching and compression as well as in-plane local rotations (see Figures S27-S29 of the SI), maximizing areas with energetically favorable stacking modes (AA'/AB' for antiparallel and AB/BA for parallel configurations). To balance this optimization, sharp domain walls of unfavorable stacking (A'B for antiparallel and AA for parallel configurations) form between the nearly commensurate regions, bearing most of the resulting stress (see Section 9 of the SI).[69] The energy penalty stemming from excess in-plane stress and unfavorable interlayer stacking is partially released by exploiting the layers' low-energy flexural modes, which elevate these domain walls and result in noticeable out-of-plane deformation. In contrast, for large twist angles (> 5°), the effect of atomic reconstruction becomes negligible, and both the upper and lower layers remain nearly flat, resulting in a lower proportion of high-efficiency thermal transport regions.

To further understand the impact of these structural changes, we investigated the relationship between high-efficiency thermal transport regions and the ITCs of twisted TMD stacks. Figure 4d illustrates a clear positive correlation between the proportion of high-efficiency thermal transport region and ITC for twisted $MoS_2$, $MoSe_2$, $WS_2$, and $WSe_2$ stacks. Notably, as the proportion of regions decreases from 100% ($\theta$ = 0°) to approximately 42% ($\theta$ = 0.7°), a significant drop in ITC is observed. This result indicates that increasing the twist angle will significantly hinder thermal transport across the twisted interface by diminishing the size of high-efficiency thermal transport regions.

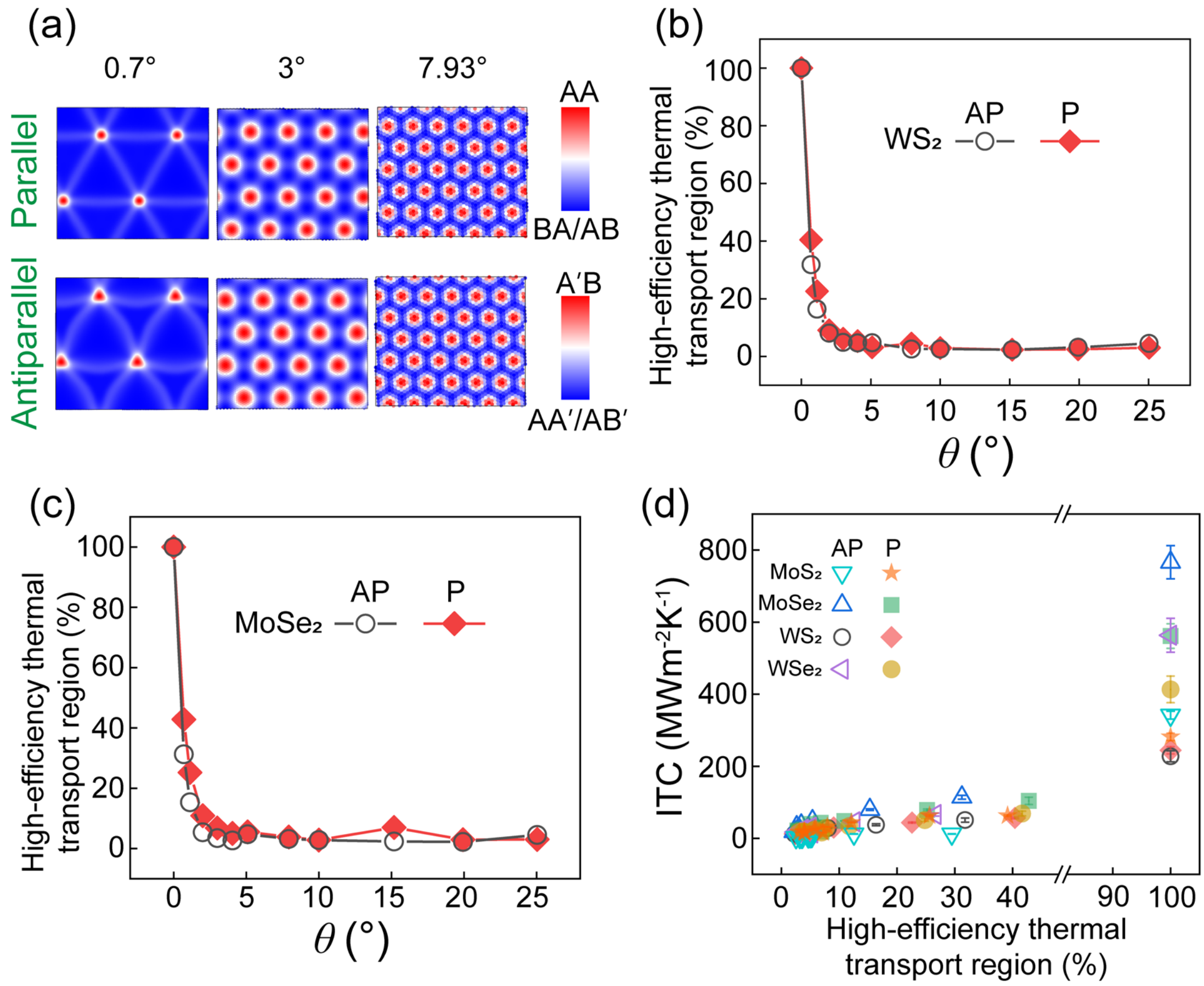

Figure 4. Relationship between high-efficiency thermal transport region and the ITCs of twisted TMD stacks. (a) Distribution of high-symmetry stacking modes in parallel and antiparallel alignment of twisted $WS_2$ bilayer under different interlayer twist angles. The proportion of high-efficiency thermal transport region of bilayer $WS_2$ (b) and $MoSe_2$ (c) as a function of the interlayer twist angle $\theta$. (d) ITCs of twisted interface as a function of the proportion of high-efficiency thermal transport regions in 16-layer parallel and antiparallel stacked TMDs.

The intricate interplay between twist angles, atomic reconstruction, and thermal transport efficiency in TMD layers makes predicting the twist-angle dependence of interlayer thermal transport in two-dimensional materials a formidable challenge. Conventionally, this task has necessitated either costly experimental measurements or computationally intensive simulations. However, our comprehensive analysis of high-efficiency thermal transport regions and their correlation with ITC provides a foundation for a more efficient approach.

To further get a deeper understanding of its underlying mechanisms, we performed phonon spectral energy density (SED)[70, 71] calculations for antiparallel twisted $WS_2$ with a twist angle of 0°, 2.9°, 9.43°, and 21.8°. The SED method analyzes atomic velocity data obtained from MD simulations to predict phonon dispersion relations, lifetimes, and thermal conductivity contributions, offering a versatile and accurate approach for thermal transport analysis in complex materials that inherently accounts for temperature-dependent anharmonicity. As demonstrated in Figure 5a, the slope of the transverse acoustic (TA) phonon branch decreases rapidly as the twist angle increases, indicating softening of acoustic phonon modes. Since the slope corresponds to the group velocity and the thermal conductivity can be approximated as $\kappa = \frac{1}{3} C_V l v_{\mathrm{g}}$,[72] where $v_{\mathrm{g}}$ is the phonon group velocity, $l$ is the phonon mean free path and $C_V$ is the heat capacity. We expect the thermal conductivity $\kappa$ share a similar twist-angle dependence to that of group velocity ($v_{\mathrm{g}}$). By extracting the group velocity and plotting it as a function of twist angle, the results (see Figure 5b) show that the relation between $v_{\mathrm{g}}$ and $\theta$ can be captured by the following expression: $v_{\mathrm{g}} = c^{\left(e^{-\sqrt{\theta}}-1\right)} + d$, where *c* and *d* are fitting parameters.

The findings above indicate that the twist-angle dependence of cross-plane thermal conductivity originates from the acoustic phonon softening caused by interfacial twisting, which is closely related to the variation in stacking mode distribution due to the atomic reconstruction within the moiré superlattices. Consequently, we propose an analytical formula to describe the relationship between interlayer twist angles and ITC, which is expressed as follows:

$$\frac{\mathrm{ITC}(\theta)}{\mathrm{ITC}_0} = \lambda^{\left(e^{-\sqrt{\theta}}-1\right)}, \lambda = a\frac{E_{\mathrm{high}}^{\mathrm{b}} - E_{\mathrm{low}}^{\mathrm{b}}}{E_{\mathrm{high}}^{\mathrm{b}}} + b \qquad (1)$$

where $\theta$ represents the misfit angle between the two twisted TMD slabs, $\mathrm{ITC}(\theta)$ and $\mathrm{ITC}_0$ represents the ITC at a twist angle of $\theta$ and 0°, respectively. $E_{\mathrm{high}}^{\mathrm{b}}$ and $E_{\mathrm{low}}^{\mathrm{b}}$ represent the binding energies of the high-symmetry stacking modes with high and low binding energy, respectively. Specifically, for the antiparallel configuration, $E_{\mathrm{high}}^{\mathrm{b}} = E_{\mathrm{AA'/AB'}}^{\mathrm{b}}$ and $E_{\mathrm{low}}^{\mathrm{b}} = E_{\mathrm{A'B}}^{\mathrm{b}}$. In contrast, for the parallel configuration, $E_{\mathrm{high}}^{\mathrm{b}} = E_{\mathrm{AB/BA}}^{\mathrm{b}}$ and $E_{\mathrm{low}}^{\mathrm{b}} = E_{\mathrm{AA}}^{\mathrm{b}}$. All the binding energies were obtained from the DFT data of our previous works.[35, 36] The dimensionless parameters $a$ and $b$ depends on the stacking type of TMD structures (antiparallel or parallel) but are independent of specific systems (see Section 10 of SI).

Figure 5c display the ln ($\mathrm{ITC}_0/\mathrm{ITC}$) as a function of $\left(1 - e^{-\sqrt{\theta}}\right)$ ln ($\lambda$) for parallel and antiparallel stacked TMDs, calculated using our general expression (eq 1) and compared with results obtained from computationally intensive NEMD simulations across a wide range of twist angles (0° ~ 25°). The close match between the theoretical results and NEMD simulations validates the accuracy and reliability of our approach. Notably, eq 1 successfully captures key features observed in the above analysis: the sharp drop in thermal conductivity at small twist angles (0° ~ 5°), where atomic reconstruction plays a significant role, and the relatively low conductivity at larger angles, where reconstruction effects become negligible. This concordance demonstrates that the proposed expression effectively translates complex physical processes (including moiré pattern formation, atomic reconstruction, the resulting changes in high-efficiency thermal transport regions and phonon softening) into a concise mathematical model. Furthermore, the applicability of this general expression (eq 1) extends beyond 2H-TMD to other 2D materials such as graphene and *h*-BN systems (see Figure 5c), and heterogeneous twisted 2H-TMDs structures (see Section 10 of SI), underscoring its versatility and potential broad impact in the field of 2D material thermal transport. However, it should be noted that our model does have certain limitations. First, it considers interlayer twisting as the sole physical factor influencing thermal transport. The introduction of external strain and defects may render our model inadequate. Second, we have focused exclusively on 2H-TMD systems ($MoS_2$, $MoSe_2$, $WS_2$ and $WSe_2$), and the validity of the model for 1T-TMD and 3R-TMD systems still needs to be assessed. As accurate interlayer force fields for these structures are currently lacking, we will address this as a direction for future work.

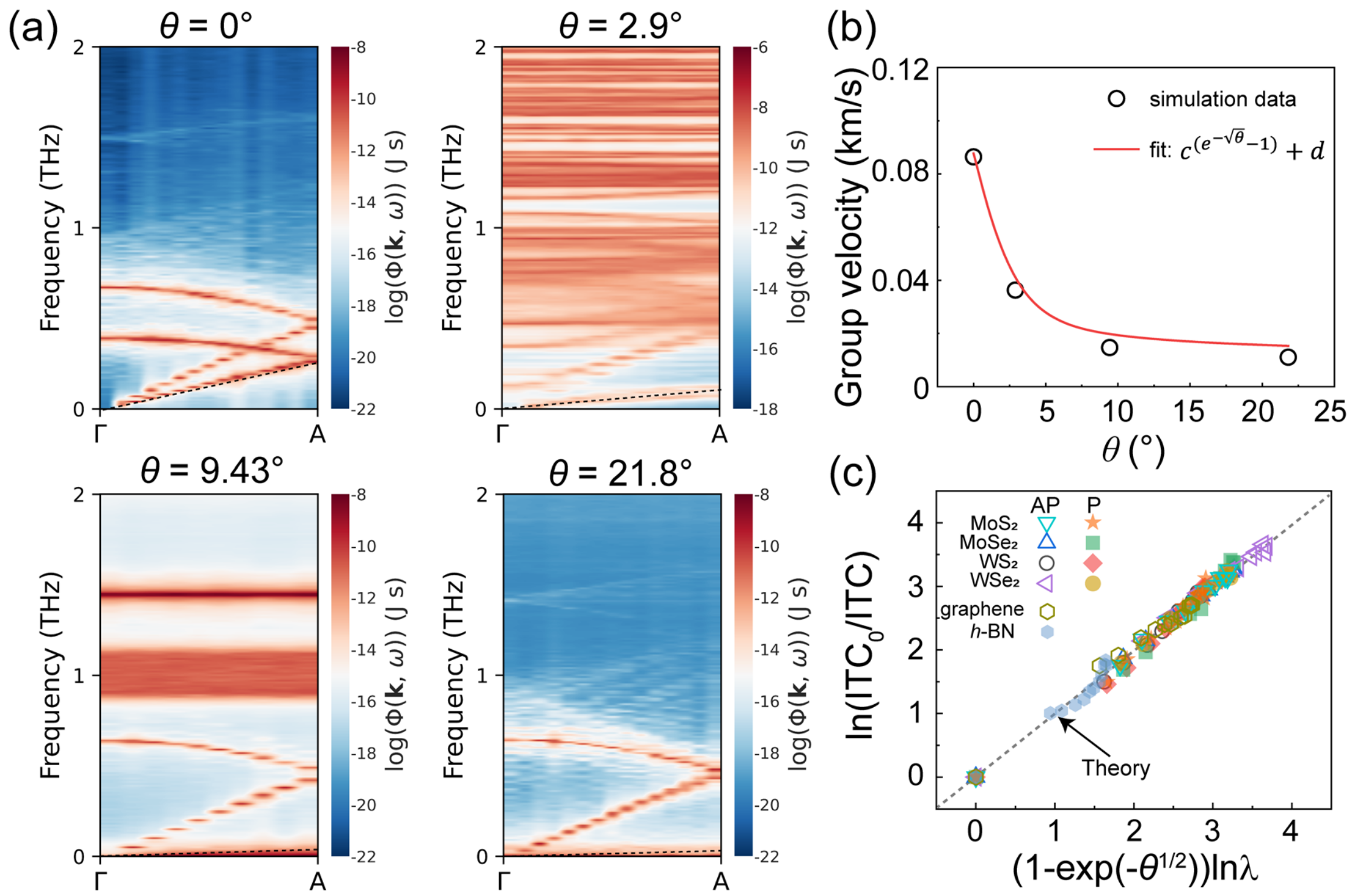

Figure 5. (a) Phonon spectral energy density distribution of twisted antiparallel $WS_2$ for twist angles of 0°, 2.9°, 9.43°, and 21.8° at 300 K. The phonon frequency distribution is presented as a function of wave number in the Γ(0, 0, 0) → A(0, 0, 0.5) path. The black dashed line represents the slope of the transverse acoustic phonon mode. (b) Dependence of group velocity ($v_g$) on twist angle for twisted $WS_2$. The black circles represent the simulation data, while the red solid line indicates the theoretical fit. The adopted theoretical formula is $v_g = c^{\left(e^{-\sqrt{\theta}}-1\right)} + d$, with the fitted parameters $c$ and $d$ are 1.08 and -0.91, respectively. (c) Theoretical prediction of the twist-angle dependence on ITC for TMD, graphene, and $h$-BN structures. Comparison between the theoretical results (dashed lines, calculated using eq 1) and the calculated ITCs obtained from NEMD simulations (hollow and solid symbols) for antiparallel and parallel TMD, graphene, and $h$-BN configurations at various twist angles. The fitted values of dimensionless parameters $a$ and $b$ are presented in Tables S7-S8 in Section 10 of the SI.

## CONCLUSIONS

In summary, our study establishes a crucial link between interlayer thermal transport in twisted TMD structures and their moiré superlattices. We demonstrate that the twist-angle dependence of cross-plane thermal conductivity is primarily driven by changes in high-efficiency thermal transport regions and softening of transverse acoustic phonon modes, both significantly influenced by atomic reconstruction at twisted interfaces. Based on these insights, we propose a concise theoretical formula that accurately describes the twist-angle-dependent thermal conductance of twisted interfaces. Our model, validated against NEMD simulations for various 2H-TMD materials and applicable to other 2D systems, offers a powerful alternative to resource-intensive experiments and simulations. This theoretical framework not only deepens our understanding of thermal transport in twisted vdW materials but also provides an invaluable predictive tool for future research and applications. By enabling rapid evaluation of thermal properties across different twist angles and material systems, our model accelerates advancements in 2D material thermal management, paving the way for innovative device designs and optimizations.

## METHODS

**Periodic Commensurate Structures for Twisted TMDs.** For the bilayer twisted TMD system, we first defined the primitive lattice vector of the bottom layer as $\boldsymbol{a}_1 = a(1,0)$ and $\boldsymbol{a}_2 = a\left(1/2, \sqrt{3}/2\right)$, where $a$ is the lattice constant of monolayer TMD. Then the lattice vector of the bottom and upper layers can be given by $\boldsymbol{L}_1 = i_1\boldsymbol{a}_1 + i_2\boldsymbol{a}_2$ and $\boldsymbol{L}_2 = j_1\boldsymbol{a}_1 + j_2\boldsymbol{a}_2$, respectively. Thus, the exact superlattice period with angle $\theta$ between two lattice vectors $\boldsymbol{L}_1$ and $\boldsymbol{L}_2$ is then given by:[28, 29]

$$L = |i_1\boldsymbol{a}_1 + i_2\boldsymbol{a}_2| = a\sqrt{i_1^2 + i_2^2 + i_1 i_2} = |j_1\boldsymbol{a}_1 + j_2\boldsymbol{a}_2| = a\sqrt{j_1^2 + j_2^2 + j_1 j_2} = \frac{|i_1 - i_2|a}{2\sin(\theta/2)} \tag{2}$$

It is important to note that the lattice structure maintains rigorous periodicity only at certain specific twist angles, and the parameters used to model 16-layers twisted TMD systems with various misfit angles are shown in Table S4 of SI Section 2.

Then, we can model the periodic supercells by using the parameters in Table S4. The initial lattice constant of $MoS_2$, $MoSe_2$, $WS_2$, and $WSe_2$ were taken as 3.128 Å, 3.311 Å, 3.128 Å, and 3.288 Å (the equilibrium value obtained using the SW intralayer potential for monolayer TMD layer). The initial interlayer distance across the layered stack was set as 6.2 Å, 6.6 Å, 6.2 Å, and 6.6 Å for $MoS_2$, $MoSe_2$, $WS_2$, and $WSe_2$, respectively. The intralayer interactions within each TMD layer employed the SW potential, and the TMD/TMD interlayer interactions were modeled *via* the ILP. Periodic boundary conditions were applied in all directions.

**Simulation Protocol.** All NEMD simulations were performed adopting the following protocol in LAMMPS.[73] We utilized a fixed time step of 0.5 fs to integrate the equations of motion. First, the systems underwent an isothermal-isobaric equilibration at a temperature of 300 K and zero pressure, employing a Nosé-Hoover thermostat with a time constant of 0.25 ps for 500 ps to relax the box (see Section 2 of SI). Our recent studies demonstrate that once the system has been equilibrated for a sufficiently long time, its effect on thermal transport properties becomes negligible.[28, 29] The interlayer thermal transport simulations then were conducted by establishing a temperature gradient (see Section 2 of SI), setting $T_{\text{cold}}$ = 225 K (heat sink) and $T_{\text{hot}}$ = 375 K (heat source) at the bottom and middle layers, respectively, using the Langevin thermostat method for 750 ps. Meanwhile, all non-thermostated layers followed the microcanonical ensemble (NVE). After the steady-state was obtained, the last 500 ps of simulation data were used to calculate the cross-plane thermal conductivity of the twisted TMD system. To estimate statistical errors, ten different datasets were used, each calculated over a time interval of 50 ps.

**Calculations of Interfacial Thermal Resistance.** In NEMD simulations, the Langevin thermostat was used to control temperature in the hot reservoir and the cold reservoir, and then the heat flux $\dot{Q}$ can be generated by the heat bath, which is defined as:

$$\dot{Q} = |\mathrm{d}E/\mathrm{d}t| \tag{3}$$

where $|\mathrm{d}E/\mathrm{d}t|$ represents the energy exchange rate between the hot reservoir and the cold reservoir during the nonequilibrium steady state. Next, the $\kappa_{\text{CP}}$ can be calculated by Fourier's law:[74]

$$\kappa_{\text{CP}} = \frac{\dot{Q}/A}{\Delta T/\Delta z} \tag{4}$$

where $\Delta T/\Delta z$ is the temperature gradient along the direction of heat flux $\dot{Q}$. $A$ is the cross-section area. To investigate the thermal properties of the twisted interface, the ITR based on the Kapitza[62] resistance can be calculated as:[28, 29]

$$(N/2-3)R_{\text{A}} + R_{\theta} = [(N/2-3)d_{\text{A}} + d_{\theta}]/\kappa_{\text{CP}} \tag{5}$$

where $d_{\text{A}}$ and $d_{\theta}$ are the interlayer distance for adjacent untwisted and twisted TMD layers. $N$ is the number of layers of the entire system. $R_{\text{A}}$ and $R_{\theta}$ are the ITRs of adjacent untwisted and twisted TMD layers, respectively. For the untwisted system ($\theta$ = 0°), $R_A$ equals to $R_{\theta}$.

**Calculations of Spectral Heat Current Decomposition.** In this work, we investigated the SHC decomposition across the twisted interface located between the heat sink and heat source to quantitatively elucidate the differences in phonon thermal transport across distinct stacking regions. The interparticle spectral heat current is defined as:[29]

$$q_{i\to j}(\omega) = -\frac{2}{\omega t_{\text{sim}}} \sum_{\alpha,\beta\in\{x,y,z\}} \text{Im}\,\langle \hat{v}_i^{\alpha}(\omega) * K_{ij}^{\alpha\beta}\,\hat{v}_j^{\beta}(\omega)\rangle\,, \tag{6}$$

where $\omega$ is the phonon frequency and the $t_{\text{sim}}$ is the total simulation time. $\hat{v}_i^{\alpha}(\omega)$ and $\hat{v}_j^{\beta}(\omega)$ are the Fourier-transformed atomic velocities of atom $i$ in direction $\alpha$ and atom $j$ in direction $\beta$, respectively. The $K_{ij}^{\alpha\beta}$ represents the harmonic interatomic force constant matrix. The $\langle\cdot\rangle$ represents the ensemble average, which is assumed to be equal to the time average due to ergodicity. The atom *i* is moved to the ±*x*, ±*y*, and ±*z* directions with a small displacement value Δ = 0.01 Å (keeping all the other atoms at their relaxed positions), respectively. After each atom has been displaced in three directions, the force constant can be calculated as follows:

$$K_{ij}^{\alpha\beta} = \frac{F_j^{\beta-} - F_j^{\beta+}}{2\Delta}, \tag{7}$$

where $F_j^{\beta-}$ and $F_j^{\beta+}$ denote force on atom $j$ in the *β*-direction, when *i* atoms are displaced to the -*α* and +*α* directions, respectively. The spectral phonon thermal conductance $G(\omega)$ can be obtained by summing over all pairs of atoms in the defined group by the temperature difference ($\Delta T$) and cross-sectional area (*A*):

$$G(\omega) = \frac{1}{A\Delta T} \sum_{i\in L} \sum_{j\in R} q_{i\to j}(\omega). \tag{8}$$

Two adjacent regions in the middle of the model are selected for the "$L$" and "$R$" groups. In our calculations, the "$L$" groups are set to the whole bottom layer of the twisted interface, and the "$R$" groups refer to the specific stacking regions in the top layer of the twisted interface. For parallel (antiparallel) configurations, we individually computed $G(\omega)$ for four stacking regions: AA, AB, BA, and DW (AA′, AB′, A′B, and DW), and obtained the corresponding ITC by summing the integrals. Finally, we compared the normalized ITC results to observe the thermal transport properties across different stacking regions. For different stacking regions, we used the same number of atoms for the "$R$" groups to compute the $G(\omega)$. Both $\Delta T$ and *A* use the temperature difference and cross-sectional area of the whole twisted interface.

**Calculations of Phonon Spectral Energy Density.** Here, we conduct calculations of the phonon spectral energy density[70, 71] to investigate the phonon dispersion characteristics of twisted $WS_2$. The phonon spectral energy density, which depends on wave vector (***k***) and frequency (*ω*), can be calculated using the following formula in MD simulations:

$$\Phi(\boldsymbol{k},\omega) = \frac{1}{4\pi\tau_0 N} \sum_{\alpha}^{\{x,y,z\}} \sum_{b}^{B} m_{\mathrm{b}} \times \left| \int_0^{\tau_0} \sum_{n_{x,y,z}}^{N} v_\alpha \binom{n_{x,y,z}}{b}; t\right) \times \exp\left[i\boldsymbol{k}\cdot \boldsymbol{r}\binom{n_{x,y,z}}{0} - i\omega t\right] \mathrm{d}t \right|^2 \tag{9}$$

where $\tau_0$ is the total simulation time, *N* is the number of unit cells, *b* is the atom label in a given unit cell *n*, $m_b$ is the mass of atom b in the unit cell, $n_{x,y,z}$ is the index number of unit cells along the *x*, *y*, and *z* directions, $v_\alpha\left(\begin{smallmatrix} n_{x,y,z} \\ b \end{smallmatrix}; t\right)$ denotes the velocity of atom *b* of the *n*-th unit cell along the *α* direction at time *t*, and $\boldsymbol{r}\binom{n_{x,y,z}}{0}$ is the equilibrium position of unit cell *n*.

## ASSOCIATED CONTENT

### Data Availability Statement

The source code for spectral heat current decomposition calculations is available at https://github.com/Tingliangstu/New-Version-Spectral-decomposition-python-tools. The source code for spectral energy density calculations is available at https://github.com/Tingliangstu/pySED. The anisotropic interlayer potential for TMD systems is available at https://docs.lammps.org/pair_ilp_tmd.html.

### Supporting Information

The Supporting Information is available free of charge on the ACS Publications website at DOI: XXXXXXXX, which contains the following sections: Refined ILP Parameters for TMD Heterostructures Containing $WSe_2$, Methodology, Convergence Tests, Thermal Transport Properties of $MoS_2$ and $WSe_2$, Distribution of Interlayer Potential Energy in Twisted TMD Interfaces, Effect of Interlayer Interaction Strength on Cross-Plane Thermal Conductivity, Calculation of Spectral Heat Current Decomposition, Definition of High-Efficiency Thermal Transport Region, Characterization and Effect of Atomic Reconstruction, and Theoretical Description of Interfacial Thermal Conductance in Twisted TMD Interfaces.

### Corresponding Author

*E-mail: w.g.ouyang@whu.edu.cn

### Notes

The authors declare no competing financial interest.

## ACKNOWLEDGMENTS

The authors acknowledge support from National Natural Science Foundation of China (Nos. 12472099, U2441207 and 12102307), the National Key R&D Project from the Ministry of Science and Technology of China (Grant No. 2022YFA1203100), the Fundamental Research Funds for the Central Universities (No. 600460100), the RGC GRF (Grant No. 14220022) and the Research Grants Council of Hong Kong (Grant No. AoE/P-701/20). Computations were conducted at the Supercomputing Center of Wuhan University, the National Supercomputer TianHe-1(A) Center in Tianjin, and Computing Center in Xi'an.